\documentclass[preprint2]{aastex}

\shorttitle{The nature and composition of Jupiter's building blocks}
\shortauthors{Mousis et al.}

\usepackage[T1]{fontenc}
\usepackage{graphicx}
\usepackage{etex}
\usepackage {amsmath}
\usepackage[squaren, Gray, cdot]{SIunits}
\usepackage{color}
\def\Mearth{M_\oplus}
\usepackage{lineno}

\begin{document}


\title{The nature and composition of Jupiter's building blocks derived from the water abundance measurements by the Juno spacecraft}


\author{Olivier Mousis\altaffilmark{1}, Jonathan I. Lunine\altaffilmark{2}, and Artyom Aguichine\altaffilmark{1}}


\altaffiltext{1}{Aix Marseille Univ, CNRS, CNES, LAM, Marseille, France {\tt olivier.mousis@lam.fr}}
\altaffiltext{2}{Department of Astronomy, Cornell University, Ithaca, NY 14853, USA}



\begin{abstract}
The microwave radiometer aboard the Juno spacecraft provided a measurement of the water abundance found to range between $\sim$1 and 5.1 times the protosolar abundance of oxygen in the near-equatorial region of Jupiter. Here, we aim to combine this up-to-date oxygen determination, which is likely to be more representative of the bulk abundance than the  Galileo probe subsolar value, with the other known measurements of elemental abundances in Jupiter, to derive the formation conditions and initial composition of the building blocks agglomerated by the growing planet, and that determine the heavy element composition of its envelope. We investigate several cases of icy solids formation in the protosolar nebula, from the condensation of pure ices to the crystallization of mixtures of pure condensates and clathrates in various proportions. Each of these cases correspond to a distinct solid composition whose amount is adjusted in the envelope of Jupiter to match the O abundance measured by Juno. The volatile enrichments can be matched by a wide range of planetesimal compositions, from solids exclusively formed from pure condensates or from nearly exclusively clathrates, the latter case providing a slightly better fit. The total mass of volatiles needed in the envelope of Jupiter to match the observed enrichments is within the $\sim$4.3--39 $\Mearth$ range, depending on the crystallization scenario considered in the protosolar nebula. A wide range of masses of heavy elements derived from our fits is found compatible with the envelope's metallicity calculated from current interior models.
\end{abstract}


\section{Introduction}

A long awaited measurement at Jupiter is the determination of its bulk oxygen abundance. This measurement is key because it provides clues on the formation conditions of the giant planet in the protosolar nebula (PSN) \citep{Ow99,Ga01,Mo09,Mo19}. The {\it in situ} measurements made by the Galileo probe up to 22 bar in Jupiter's atmosphere showed that the O abundance was subsolar ($\sim$0.46 times the protosolar O abundance; \cite{Wo04}). This result was mostly attributed to the dynamics of the region within which the Galileo probe descended \citep{Or98}, but alternative interpretations suggested that the measurement may correspond to the bulk composition of the planet \citep{Mo12}. However, the findings of the Galileo Probe were not conclusive because the data showed that the abundance of water was still increasing with progressing depth, until loss of signal. Recently, the microwave radiometer ({MWR)} aboard the Juno spacecraft provided a measurement of the water abundance in the equatorial region of Jupiter, from 0 to 4 degrees of north latitude, and in the 0.7--30 bar pressure domain. The water abundance in this region of Jupiter was found to range between $\sim$1 and 5 times the protosolar abundance of oxygen, a value substantially higher than the Galileo probe determination, and compatible with the presence of a moist adiabatic temperature profile at the water condensation level \citep{Li20}.

Here, we aim to combine this up-to-date oxygen determination, which is likely to be more representative of the bulk abundance than the Galileo probe value, with the other known measurements of elemental abundances in Jupiter, to derive the formation conditions and initial composition of the building blocks acquired by the growing planet, and that determined the composition of its envelope, in the framework of the core accretion model \citep{Po96,In03,Po20}. To do so, we also use the recent Juno N abundance measurement that is found correlated with the O abundance in Jupiter's envelope \citep{Li20}, as well as the {\it in situ} determinations of Ar, Kr, Xe, C, N, and S by the Galileo probe \citep{Ma00,Wo04}, and the P abundance inferred from infrared observations by the Cassini spacecraft \citep{Fl09} (see Table \ref{tab1}).

We investigate several cases of icy solids formation in the PSN, from the condensation of pure ices to the crystallization of mixtures of pure condensates and clathrates in various proportions. Each of these cases corresponds to a distinct solid composition whose amount is adjusted in the envelope of Jupiter to match the O abundance measured by Juno. The volatile enrichments can be matched by a wide range of planetesimal compositions, from solids exclusively formed from pure condensates or from nearly pure clathrates, these latter providing the best fits. Estimates of the corresponding amounts of heavy elements in the envelope are also provided and compared to interior models derived from Juno observations.


\begin{table*}
\begin{center}
\caption[]{Elemental abundances and ratios to protosolar values in the upper troposphere of Jupiter}
\label{tab1}
\begin{tabular}{lccccc}
\hline
\hline
\noalign{\smallskip}
Elements			& Elemental abundances 					& Jupiter					& References\\
 				&  									& /Protosolar$^1$				& \\
\hline
\noalign{\smallskip}
Ar/H				& ($9.10 \pm 1.80) \times 10^{-6}$			&	$3.30 \pm 0.65$		& \cite{Ma00}					\\
Kr/H				& ($4.65 \pm 0.85) \times 10^{-9}$			&	$2.38 \pm 0.44$		& \cite{Ma00}					\\
Xe/H	 			& ($4.45 \pm 0.85) \times 10^{-10}$			&	$2.33 \pm 0.45$		& \cite{Ma00}					\\
C/H 				& ($1.19 \pm 0.29) \times 10^{-3}$			&	$4.01 \pm 0.97$		& \cite{Wo04}					\\
N/H 				& ($2.04 \pm 0.13) \times 10^{-4}$			& 	$2.75 \pm 0.17$		& \cite{Bo17,Li17,Li20}			\\
O/H				& ($2.45 \pm 0.80) \times 10^{-4}$			&  	$0.46 \pm 0.15$ 		& \cite{Wo04}					\\
 				& ($1.45^{+1.27}_{-0.93}) \times 10^{-3}$		&	$2.69^{+2.37}_{-1.72}$	& \cite{Li20}					\\
S/H 				& ($4.45 \pm 1.05) \times 10^{-5}$			&	$3.08 \pm 0.73$		& \cite{Wo04}					\\
P/H 				& ($1.08 \pm 0.06) \times 10^{-6}$			&	$3.82 \pm 0.20$		& \cite{Fl09}					\\			
\hline
\end{tabular}
\end{center}
\end{table*}

\section{Building blocks composition model}

In our model, the volatile phase incorporated in planetesimals is composed of a mixture of pure ices, stoichiometric hydrates (such as NH$_3$-H$_2$O hydrate), and clathrates that crystallized in the form of microscopic grains at various temperatures in the outer part of the disk. Solids accreted by the growing Jupiter are assumed to have preserved the volatile budget acquired during condensation of the microscopic ices from which they assembled, regardless of the way they evolved in the PSN (accretion of pebbles or large planetesimals,  products of collisions, etc).

We assume that the PSN is uniformly filled with H$_2$O, CO, CO$_2$, CH$_3$OH, CH$_4$, N$_2$, NH$_3$, H$_2$S, and PH$_3$. These molecules are considered to be the dominant volatile species in the PSN, assuming protosolar abundances for O, C, N, S, and P \citep{As09}. Half of the sulfur is assumed to be in the form of H$_2$S with the other half forming refractory sulfide components \citep{Pa05}, and all C forms CO, CO$_2$, CH$_3$OH and CH$_4$, with the remaining O going into H$_2$O. We have set CO:CO$_2$:CH$_3$OH:CH$_4$ = 10:30:1.67:1 in the gas phase of the disk. The CO:CO$_2$ ratio comes from ROSINA observations of the coma of comet 67P/Churyumov-Gerasimenko (hereafter 67P/C-G) \citep{La19}. The CO:CH$_3$OH:CH$_4$ ratio is consistent with the production rates measured in the southern hemisphere of 67P/C-G in October 2014 by the ROSINA instrument \citep{Le15}. We also assume N$_2$:NH$_3$ = 1:1 in the nebula gas phase \citep{Fe00}.

The process of volatile trapping in grains crystallized in the outer regions of the PSN is determined by using the equilibrium curve of NH$_3$ hydrate, those of the various clathrates and pure condensates, as well as the thermodynamic path detailing the evolution of temperature and pressure at 5.2 AU (i.e., the current location of Jupiter) in the protoplanetary disk. We refer the reader to the work of \cite{Mo09} for a full description of the calculation method regarding the composition of the solids crystallized in the PSN. The equilibrium curves of hydrates and clathrates derive from the compilation of published experimental work by \cite{Lu85}, in which data are available at relatively low temperatures and pressures. The equilibrium curves of pure condensates used in our calculations derive from the compilation of laboratory data from \cite{Li02}. The fits of the equilibrium data are taken from \cite{He04} and \cite{Mo08}.

The disk model employed is the one described in \cite{Ag20} and \cite{Mo20}, to which the reader is referred for details. In a few words, our time-dependent PSN model is governed by the following differential equation \citep{Ly74}:

\begin{eqnarray}
\frac{\partial \Sigma_{\mathrm{g}}}{\partial t} = \frac{3}{r} \frac{\partial}{\partial r} \left[ r^{1/2} \frac{\partial}{\partial r} \left( r^{1/2} \Sigma_{\mathrm{g}} \nu \right)\right].
\label{eqofmotion}
\end{eqnarray}

\noindent This equation describes the time evolution of a viscous accretion disk of surface density $\Sigma_{\mathrm{g}}$ of viscosity $\nu$, assuming hydrostatic equilibrium in the $z$ direction. The viscosity $\nu$ is calculated in the framework of the $\alpha$-formalism \citep{Sh73}. {In our approach, the midplane temperature $T$ of the disk is expressed as the sum of viscous heating and star background irradiation $T_{\mathrm{amb}}$= 10 K: 

\begin{equation}
T^4 = \frac{1}{2\sigma_{\mathrm{SB}}} \left( \frac{3}{8}\tau_\mathrm{R} + \frac{1}{2 \tau_\mathrm{P}} \right) \Sigma_\mathrm{g} \nu \Omega_\mathrm{K}^2 + T^4_{\mathrm{amb}},
\label{eq:temp}
\end{equation}

\noindent where $\sigma_{\mathrm{sb}}$ is the Stefan-Boltzmann constant, $\Omega_\mathrm{K}$ is the keplerian frequency, $\tau_\mathrm{R}$ and $\tau_\mathrm{P}$ are the Rosseland and Planck optical depth, respectively (see \cite{Ag20} and \cite{Mo20} for details). Only viscous heating is considered as an energy source in our model, allowing the disk temperature to decrease down to $\sim$10 K in the outer PSN. Here viscous means eddy viscosity, namely that provided by eddies which transport mass and heat by bulk motions, since molecular viscosity is very small \citep{St90}. This corresponds to the case where the outer parts of the disk are protected from solar irradiation by shadowing effect of the inner disk parts.} The evolution of the disk starts with an initial profile given by $\Sigma_{\mathrm{g}} \nu \propto \exp \left(-r^{2-p}\right)$, with $p=\frac{3}{2}$ for an early disk \citep{Ly74}. In our computations, the initial disk mass is fixed to 0.1~M$_\odot$. The computational box is set equal to 500 AU, allowing 99\% of the disk mass to be encapsulated within $\sim$100 AU. The disk initial mass accretion rate onto the Sun is set to $10^{-7.6}$~M$_\odot$~yr$^{-1}$ \citep{Ha98}, and the viscosity parameter $\alpha$ is fixed equal to 3~$\times$~10$^{-3}$, which is well within the 10$^{-4}$--10$^{-2}$ range commonly adopted for PSN models \citep{Mo20}. 

{In the following, we assume that condensate or clathrate grains quickly decouple from gas, due to growth and planetesimal formation. The enrichment of species $X$ with respect to its protosolar abundance is then given by:

\begin{equation}
\frac{X/H_2(t)}{X/H_2|_\odot} = \frac{\Sigma_{\mathrm{g}} (R,t_X)}{\Sigma_{\mathrm{g}} (R,t)},
\label{enri}
\end{equation}

\noindent where if $t_X$ is the time at which species $X$ is trapped or condensed at given distance $r$ in the PSN, This relation is valid for the determination of the volatile enrichments in Jupiter's feeding zone, and subsequently in its envelope. A set of volatile enrichments is simultaneously calculated in Jupiter when Eq. \ref{enri} is fitted to the abundance measurement of a given element in the envelope.}

\begin{figure*}[h]
\resizebox{\hsize}{!}{\includegraphics[angle=0]{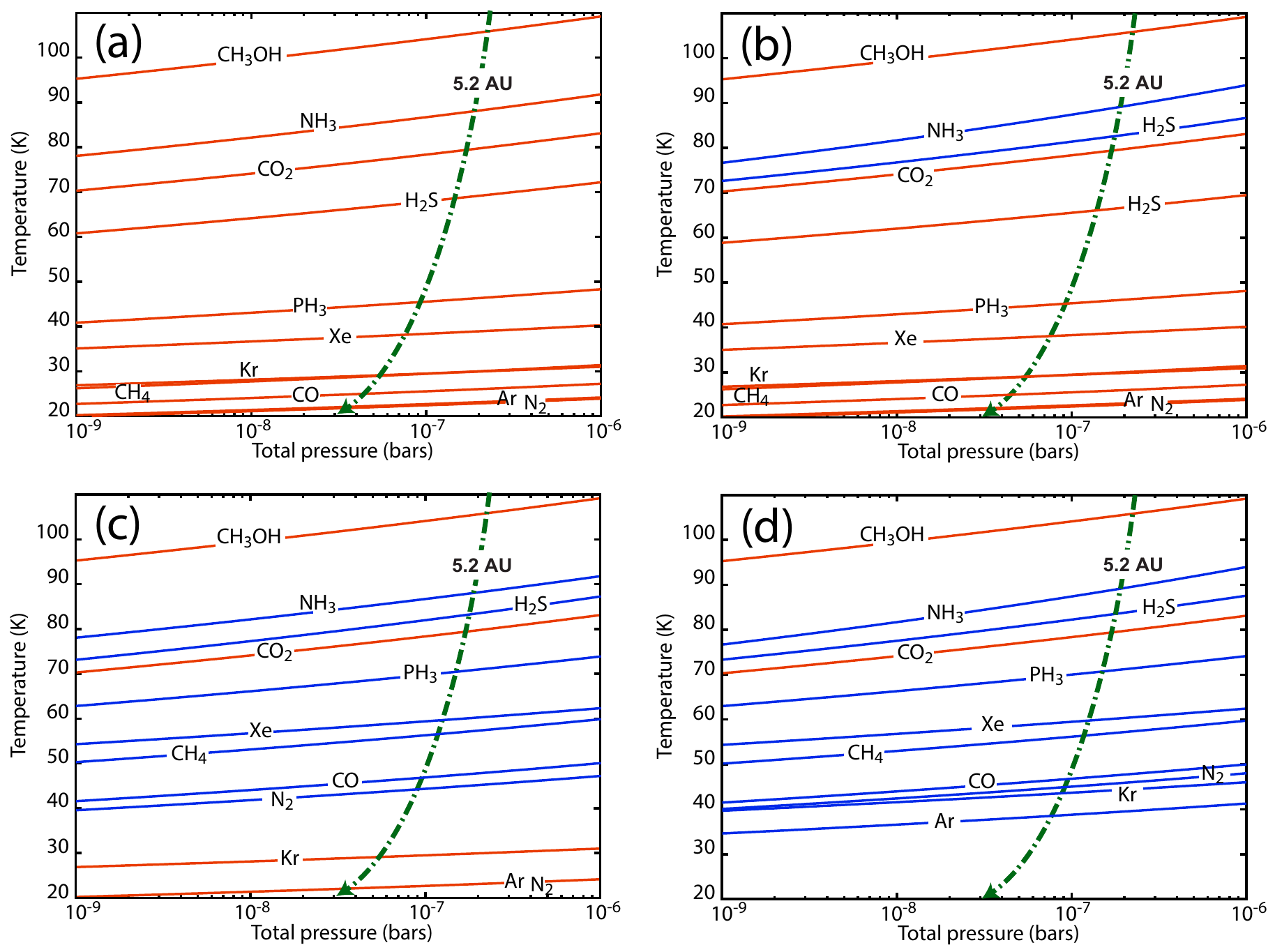}}
\caption{ Formation conditions of icy solids in the PSN defined by the intersection between various equilibrium curves and the disk cooling curve at 5.2 AU.  {Arrows pointing down in the cooling curves indicate the direction of time evolution.} Species remain in the gas phase above the equilibrium curves. Below, they are trapped as clathrates or simply condense. Panel (a): equilibrium curves of pure condensates (red lines). Abundances of various elements are protosolar \citep{As09}, with molecular mixing ratios given in the text. Panel (b): same gas phase conditions as in panel (a) but with the assumption that all available H$_2$O is used to form NH$_3$ hydrate (NH$_3$-H$_2$O) and H$_2$S clathrate (H$_2$S-5.75H$_2$O) (blue lines) in the PSN. Remaining species form pure condensates (red lines). Panel (c): same as panel (b), except that O/H$_2$ is 1.5 $\times$ (O/H$_2$)$_\sun$ in the PSN, implying that the number of entrapped species is more important. Panel (d): same as panels (b) and (c), except that O/H$_2$ is 2 $\times$ (O/H$_2$)$_\sun$ in the PSN, implying that all species potentially forming clathrates become entrapped.}
\label{fig1}
\end{figure*}

Figure \ref{fig1} represents the condensation sequence of the different volatiles at 5.2 AU in the PSN, for various initial H$_2$O abundances. The domain of stability of each ice considered is the region located below its corresponding equilibrium curve. In all panels, it is assumed that CO:CO$_2$:CH$_3$OH:CH$_4$~=~10:30:1.67:1, H$_2$S/H$_2$ = 0.5 $\times$ (S/H$_2$)$_\sun$, and N$_2$:NH$_3$ = 1:1 in the PSN gas phase, based on the aforementioned assumptions. Four cases are investigated. Case 1 (panel (a)) corresponds to the case where the abundances of all elements are protosolar \citep{As09}, assuming volatile species form only pure condensates in the PSN. Case 2 (panel (b)) corresponds to the same assumptions as in case 1 regarding the PSN gas phase composition, but here it is assumed that all available water is used to form hydrates in the PSN. In this situation, NH$_3$-H$_2$O hydrate forms first with decreasing disk temperature, and then almost 75\% of available H$_2$S is trapped in clathrates. The remaining species subsequently form pure condensates at lower disk temperatures. Note that CO$_2$ and CH$_3$OH are both represented in their pure ice forms in all our condensation sequences. Pure CO$_2$ is the only condensate that crystallizes at higher temperature than its associated clathrate in this pressure range and no experimental data concerning the equilibrium curve of CH$_3$OH clathrate has been reported in the literature. Case 3 (panel (c)) corresponds to the same assumptions as in case 2, except that O/H$_2$ is 1.5 $\times$ (O/H$_2$)$_\sun$. Here, NH$_3$-H$_2$O hydrate forms, and H$_2$S, PH$_3$, Xe, CH$_4$, and CO are fully trapped in clathrates. No more than $\sim$67\% of available N$_2$ in enclathrated because of the lack of available free water, and the remaining volatile species form pure condensates at lower disk temperatures. Case 4 (panel (d)) corresponds to the same assumptions as in case 3, except that O/H$_2$ is 2 $\times$ (O/H$_2$)$_\sun$. Here, all volatiles except CH$_3$OH and CO$_2$, form hydrates in the PSN. Note that we do not consider mixed clathrates, that is, incorporation of species at temperatures other than where their pure clathrates form. This does not significantly affect the final result in terms of comparison with Jupiter data.

\begin{figure*}[h]
\resizebox{\hsize}{!}{\includegraphics[angle=0]{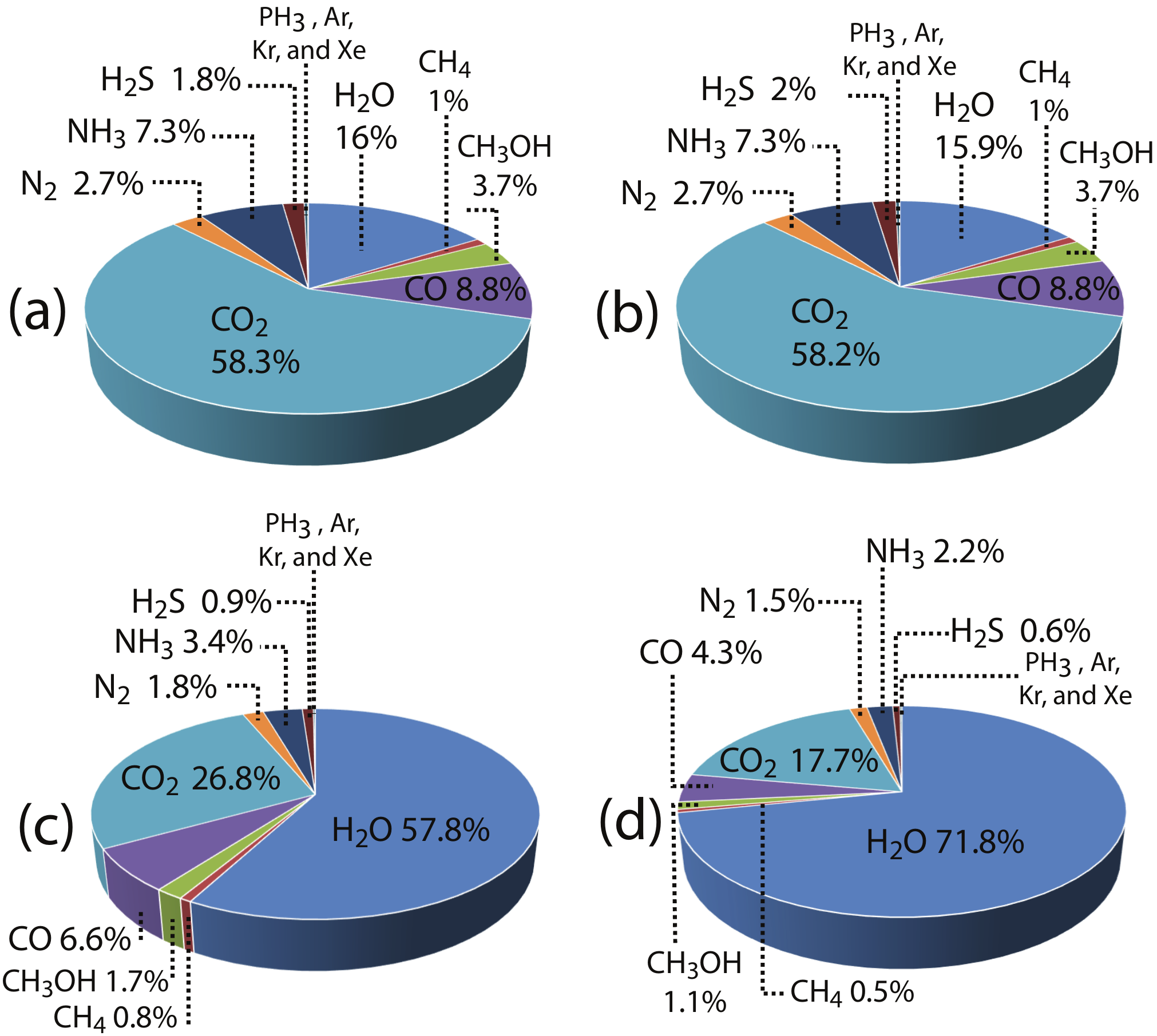}}
\caption{Composition of the volatile phase incorporated in solids in the cases the volatile part of the building blocks is formed from pure condensates only (panel (a)), a mixture of pure condensates, NH$_3$ hydrate and various clathrates (panels (b) and (c)), and from NH$_3$ hydrate and clathrates only (panel (d)). Initial gas phase conditions in panels (a)--(d) are those depicted in the text and given in Fig. \ref{fig1}.}
\label{fig2}
\end{figure*}

Figure \ref{fig2} shows the composition of the volatile phase embedded in solids formed in the PSN resulting from our model. Calculations represented in panels (a--d) follow the cases 1--4 depicted above, respectively. CO$_2$ is by far the main volatile species in panels (a) and (b) while H$_2$O dominates in panels (c) and (d). The high abundance of H$_2$O in those panels results from the choice of a larger O abundance in the PSN. In panels (b--d), the icy part of planetary building blocks is essentially made of a mixture of pure condensates and clathrates in proportions fixed by the PSN oxygen abundance.

\section{Results}

\begin{figure*}[h]
\resizebox{\hsize}{!}{\includegraphics[angle=0]{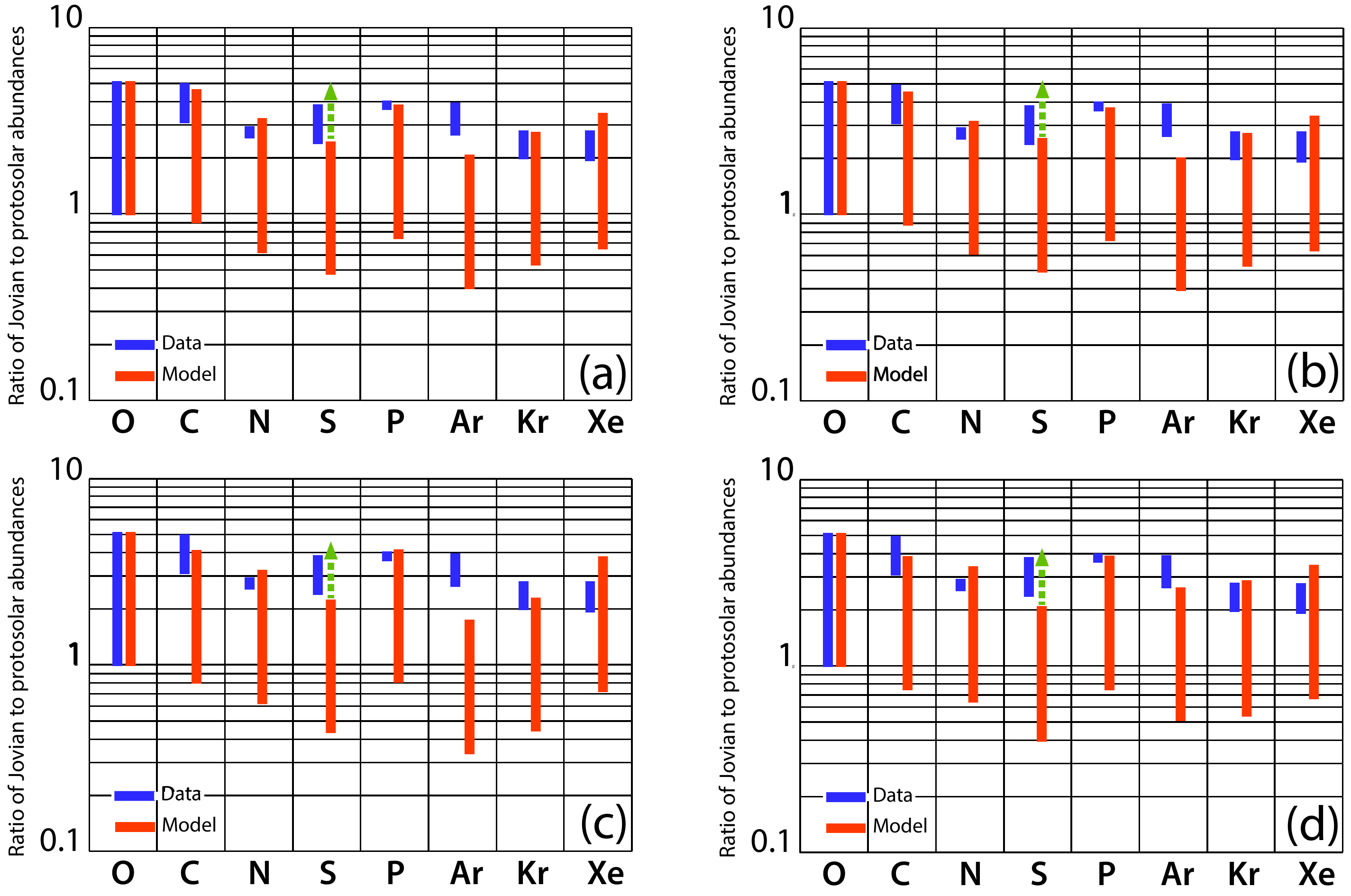}}
\caption{Ratio of Jovian to protosolar abundances in the cases the volatile part of the building blocks is formed from pure condensates only (panel (a)), a mixture of pure condensates, NH$_3$ hydrate and various clathrates (panels (b) and (c)), and from NH$_3$ hydrate and clathrates only (panel (d)). Initial gas phase conditions in panels (a)--(d) are those given in Fig. \ref{fig1}. Blue and red bars correspond to observations and model, respectively. Initial gas phase conditions in panels (a)--(d) are those depicted in the text and given in Fig. \ref{fig1}. The green arrow pointing up illustrates that fact that the calculated S enrichment is potentially higher if one assumes H$_2$S/H$_2$ $>$ 0.5 $\times$ (S/H$_2$)$_\sun$ in the PSN gas phase.}
\label{fig3}
\end{figure*}

 Figure \ref{fig3} represents the fits of the volatile enrichments (relative to protosolar abundances) observed in Jupiter, and summarized in Table \ref{tab1}. Panels (a--d) are calculated following the conditions depicted in cases 1--4, respectively. All panels have been fitted to the oxygen abundance measured by Juno in Jupiter's atmosphere. The fits consist in reproducing the mass range of oxygen measured in Jupiter by adjusting the amount of O--bearing solids (here H$_2$O, CO, CO$_2$, and CH$_3$OH) formed in the PSN and that is injected in the envelope, in all cases of planetesimals compositions. All our calculations assume the envelope is homogeneously mixed. Panels (a--d) show that C, N, P, Kr, and Xe elemental enrichments are always fitted by our different compositions models of the building blocks accreted by Jupiter. S is fitted in panels (a--b), which both correspond to the assumption that all elements, including O, are in protosolar abundances in the initial gas phase of the PSN. Panels (c--d), which assume both supersolar O abundances and clathrate formation in Jupiter's feeding zone, present calculations of S enrichments slightly lower than the measured value. However, an increase by $\sim$6--12$\%$ of the assumed H$_2$S abundance in the PSN, this latter being poorly assessed \citep{Pa05}, or the delivery to Jupiter of other S--bearing solids such as SO$_2$ in the volatile phase or FeS in the mineral phase, would easily enable the matching between the measured and calculated enrichments. The Ar enrichment is only marginally fitted in panel (d), which corresponds to case 4 (full clathration and O/H$_2$ = 2 $\times$ (O/H$_2$)$_\sun$ in the PSN). Only the full trapping of Ar in clathrates at higher temperature than its condensation in the PSN (see Fig. \ref{fig1}) allows the matching of the observed enrichment. 
 
The fitted enrichments also translate into a mass range of icy solids incorporated in Jupiter's envelope following the compositions shown in Fig. \ref{fig2}. Table \ref{tab2} summarizes these mass ranges, assuming an average envelope mass of $\sim$301.8 $\Mearth$ \citep{Wa17}. The mass ranges calculated for cases 1--4 must be seen as minimums because the accreted solids should harbor a significant fraction of minerals and metals. Assuming a fraction of rocks (minerals + metals) half the total mass of condensed matter in the PSN for a gas of protosolar composition \citep{Lo03}, the mass of accreted solids should be then twice the values indicated for cases 1 and 2 in Table \ref{tab2}, giving a total mass range between 8.6 and 45.2 $\Mearth$. The same extra mass of rocks, i.e. between $\sim$4.3 and 22.6 $\Mearth$ should be added to cases 3 and 4 since the extra oxygen added to the PSN gas in our model is assumed to go only with the volatiles.
 
\begin{table*}[h]
\begin{center}
\caption[]{Mass of volatiles needed in the envelope of Jupiter to match the observed enrichments}
\label{tab2}
\begin{tabular}{lcc}
\hline
\hline
\noalign{\smallskip}
				& Mass of H$_2$O ice ($\Mearth$)			& Mass of ices	($\Mearth$) 		\\	
\hline
\noalign{\smallskip}
Case 1			&	0.4--1.9							& 4.3--22.6					\\	
Case 2			&	0.4--1.9							& 4.3--22.7 					\\	
Case 3			&	2.3--12.2							& 5.9--30.7					\\	
Case 4 			&	4.1--21.5							& 7.4--39.0					\\	
\hline
\end{tabular}\\
\end{center}
\end{table*}

\section{Conclusion}

Our work shows that a wide range of planetesimal compositions allows the fits of the volatiles enrichments observed in Jupiter's envelope, including the recent Juno O determination. Ices either formed from pure condensates or from clathrates can match the observed enrichments in a satisfying manner. We find however that the fits of the volatile enrichments in the case of planetesimals essentially formed from clathrates is slightly more compelling since it allows marginally the matching of the Ar abundance in Jupiter, contrary to the case of pure condensates. The agglomeration of planetesimals from clathrates classically requires a supersolar abundance of oxygen at their formation location \citep{Ga01,Mo09}. Such an effect in the PSN can occur at the location of the snowline, where the outward diffusion of vapor increases the local abundance of solid water \citep{St88,Cy99,Mo19}.

The mass range of volatiles needed to be injected in solid form in Jupiter's envelope is within $\sim$4.3--39 $\Mearth$, if one considers all investigated cases. These values are well within the mass range of heavy elements derived from Juno's preliminary measurements of gravitational moments \citep{Wa17}. Based on the interior model of \cite{Mi13}, \cite{Wa17} showed that the mass of heavy elements is within the $\sim$3.3--15.9 $\Mearth$ range, depending on the type of core, which can be diluted or compact. Using the model of \cite{Be13}, these authors also derive a larger amount of heavy elements in the envelope, which is within the $\sim$14.5--40 $\Mearth$ range. Even if a substantial fraction of refractory component is taken into account in our calculations of planetesimal compositions, a wide range of masses of heavy elements derived from our fits is compatible with the envelope's metallicity calculated from current interior models.

Our calculations of the volatile budget in the envelope of Jupiter assume the envelope is homogeneously mixed. This may be inaccurate, as some recent models suggest a composition gradient in the envelope \citep{Le12,De19}. However, further theoretical and experimental studies of hydrogen-helium mixtures are needed below $\sim$100 GPa to assess these hypotheses, a region where the models are critically sensitive to changes in the equation of state \citep{Wa17}. Also, one cannot exclude that the observed volatile enrichments correspond to a late planetesimal accretion which, coupled with a rapid infall of gas, could have led to a mixing of accreted material throughout the outer regions, which may explain the supersolar metallicity \citep{Po20}. However, in this scenario, our results regarding the composition and structure of the volatile phase embedded in planetesimals would remain valid.

{Note that the Juno O determination in Jupiter considered in our work is derived from a 1--$\sigma$ uncertainty. A 2--$\sigma$ uncertainty  allows a H$_2$O abundance to be as low as 0.1 times the solar abundance  \citep{Li20}, a value lower than the one ($\sim$0.5 times solar) inferred by the Galileo probe \citep{Wo04}. If the bulk O abundance in Jupiter is significantly subsolar, this implies that the planet could belong to the category of carbon-rich planets, i.e. those with C/O $\ge$ 1 in their envelopes \citep{Mo12}. This would imply that water ice was heterogeneously distributed over several AU beyond the snow line in the protosolar nebula and that the fraction of water contained in icy planetesimals was a strong function of their formation location and time. 

Interestingly, for the sake of comparison, we have also used the compilation of protosolar abundances derived from \cite{Lo09}. Despite the important deviations noted for the C/O and Ar/O ratios in \cite{Lo09}, which are +15\% and -17\% compared with \cite{As09}, respectively, the fits of the Jovian abundances with this database lead to similar conclusions. 

One should recall that the present study is based on the Juno MWR measurements performed only at the equator and that did not penetrate deeper than the $\sim$30 bar region, without using the data from the longest wavelength, 0.6 GHz channel. Future Juno measurements at other latitudes, combined with updated interior structure models, should decrease the size of the O error bar and enable our model to provide more stringent insights on Jupiter's formation conditions in the PSN. Finally, our calculations assume that the chemical composition of the disk is homogeneous in the formation region of Jupiter's building blocks beyond the snowline. This assumption does not consider the variations in the abundances of the various ices at the locations of their condensation lines \citep{Mo20,Mo21}. Future models describing the composition of the solids formed in the PSN will have to consider these physical processes.}

\acknowledgements
 O.M. acknowledges support from CNES. J.L. was supported by the Juno project through a contract with Southwest Research Institute. We thank the anonymous referee for the helpful comments, and constructive remarks on this manuscript.



\end{document}